\documentclass[aps,reprint,superscriptaddress,amsmath,amssymb,longbibliography]{revtex4-1}

\usepackage{graphicx}
\usepackage{amsmath}
\usepackage{bm}
\usepackage{float}

\usepackage{color}

\definecolor{CLBlue}{rgb}{0, .3, .6}

\begin{document}

%\preprint{}

%\title{Non-equilibrium dynamics and entropy production in the human brain}
\title{Broken detailed balance and entropy production in the human brain}


\author{Christopher W. Lynn}
\affiliation{Initiative for the Theoretical Sciences, Graduate Center, City University of New York, New York, NY 10016, USA}
\affiliation{Joseph Henry Laboratories of Physics, Princeton University, Princeton, NJ 08544, USA}
\author{Eli J. Cornblath}
\affiliation{Department of Bioengineering, School of Engineering \& Applied Science, University of Pennsylvania, Philadelphia, PA 19104, USA}
\affiliation{Department of Neuroscience, Perelman School of Medicine, University of Pennsylvania, Philadelphia, PA 19104, USA}
\author{Lia Papadopoulos}
\affiliation{Department of Physics \& Astronomy, College of Arts \& Sciences, University of Pennsylvania, Philadelphia, PA 19104, USA}
\author{Maxwell A. Bertolero}
\affiliation{Department of Neuroscience, Perelman School of Medicine, University of Pennsylvania, Philadelphia, PA 19104, USA}
\author{Danielle S. Bassett}
%\altaffiliation{To whom correspondence should be addressed.}
\affiliation{Department of Bioengineering, School of Engineering \& Applied Science, University of Pennsylvania, Philadelphia, PA 19104, USA}
\affiliation{Department of Physics \& Astronomy, College of Arts \& Sciences, University of Pennsylvania, Philadelphia, PA 19104, USA}
\affiliation{Department of Electrical \& Systems Engineering, School of Engineering \& Applied Science, University of Pennsylvania, Philadelphia, PA 19104, USA}
\affiliation{Department of Neurology, Perelman School of Medicine, University of Pennsylvania, Philadelphia, PA 19104, USA}
\affiliation{Department of Psychiatry, Perelman School of Medicine, University of Pennsylvania, Philadelphia, PA 19104, USA}
\affiliation{Santa Fe Institute, Santa Fe, NM 87501, USA}

%\thanks{}


\date{\today}

\begin{abstract}

Living systems break detailed balance at small scales, consuming energy and producing entropy in the environment in order to perform molecular and cellular functions. However, it remains unclear how broken detailed balance manifests at macroscopic scales, and how such dynamics support higher-order biological functions. Here we present a framework to quantify broken detailed balance by measuring entropy production in macroscopic systems. We apply our method to the human brain, an organ whose immense metabolic consumption drives a diverse range of cognitive functions. Using whole-brain imaging data, we demonstrate that the brain nearly obeys detailed balance when at rest, but strongly breaks detailed balance when performing physically and cognitively demanding tasks. Using a dynamic Ising model, we show that these large-scale violations of detailed balance can emerge from fine-scale asymmetries in the interactions between elements, a known feature of neural systems. Together, these results suggest that violations of detailed balance are vital for cognition, and provide a general tool for quantifying entropy production in macroscopic systems.

\end{abstract}

%\pacs{asdfasdf}
%\keywords{keywords}

\maketitle

\section{Introduction}

\begin{figure*}
\centering
\includegraphics[width = .8\textwidth]{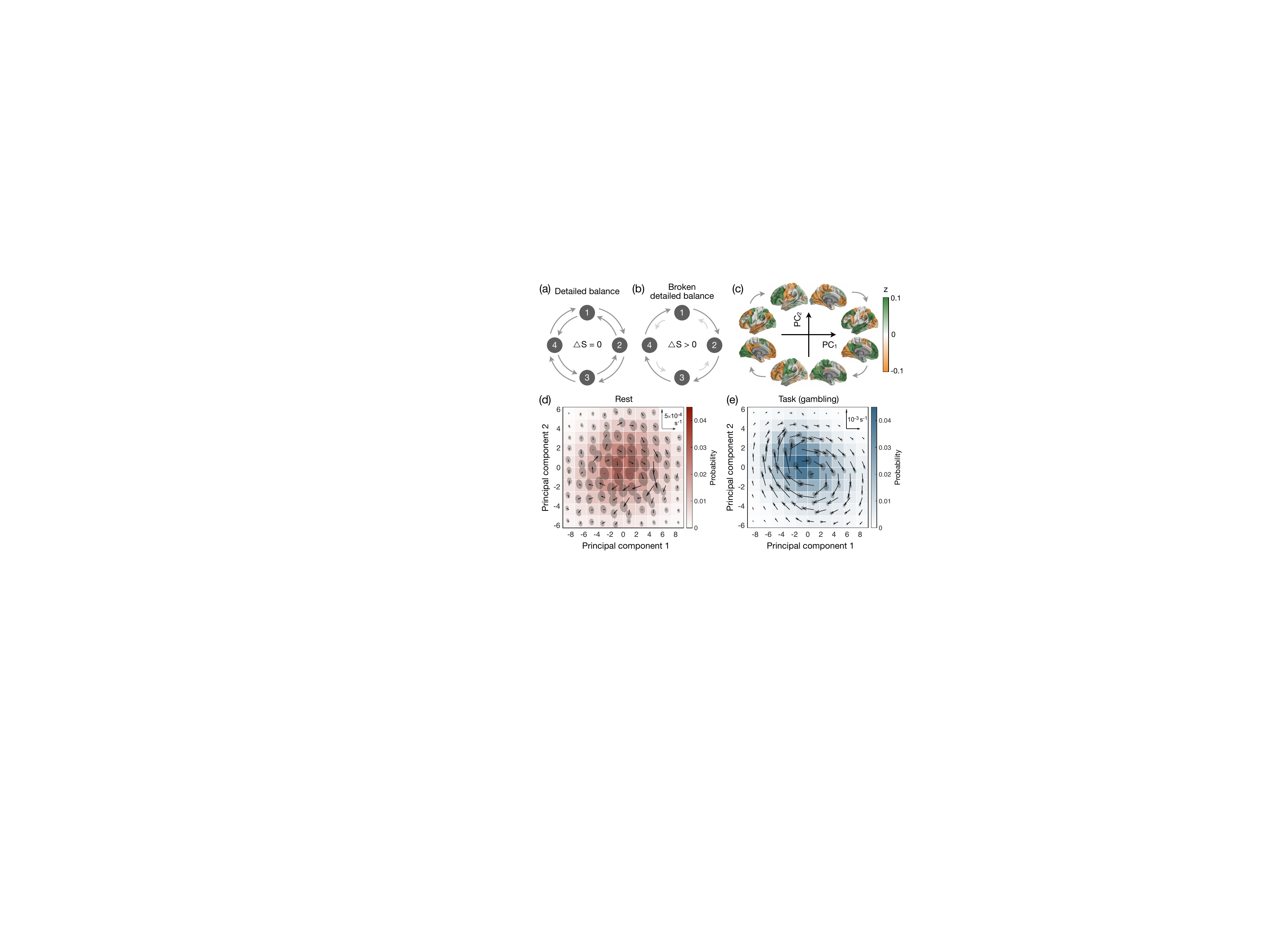} \\
\caption{\label{fluxMap} Macroscopic broken detailed balance in the brain. (a-b) A simple four-state system, with states represented as circles and transition rates as arrows. (a) During detailed balance, there are no net fluxes of transitions between states, and the system does not produce entropy. (b) Systems that break detailed balance exhibit net fluxes of transitions between states, thereby producing entropy. (c) Brain states defined by the first two principal components of the neuroimaging time-series of regional activity, computed across all time points and all subjects. Colors indicate the z-scored activation of different brain regions, ranging from high-amplitude activity (green) to low-amplitude activity (orange). Arrows represent hypothetical fluxes between states. (d-e) Probability distribution (color) and net fluxes between states (arrows) for neural dynamics at rest (d) and during a gambling task (e). In order to use the same axes in panels (d) and (e), the dynamics are projected onto the first two principal components of the combined rest and gambling time-series data. The flux scale is indicated in the upper right, and the disks represent two-standard-deviation confidence intervals that arise due to finite data (see Appendix \ref{methods}; Fig. \ref{fluxVec}).}
\end{figure*}

The functions that support life -- from processing information to generating forces and maintaining order -- require organisms to break detailed balance \cite{Schrodinger-01, Gnesotto-01}. For a system that obeys detailed balance, the fluxes of transitions between different states vanish [Fig. \ref{fluxMap}(a)]. The system ceases to produce entropy and its dynamics become reversible in time. By contrast, living systems exhibit net fluxes between states or configurations [Fig. \ref{fluxMap}(b)], thereby breaking detailed balance and establishing an arrow of time \cite{Gnesotto-01}. Critically, such broken detailed balance leads to the production of entropy, a fact first recognized by Sadi Carnot in his pioneering studies of irreversible processes \cite{Carnot-01}. At the molecular scale, enzymatic activity drives non-equilibrium processes that are crucial for intracellular transport \cite{Brangwynne-01}, high-fidelity transcription \cite{Yin-01}, and biochemical patterning \cite{Huang-01}. At the level of cells and subcellular structures, broken detailed balance enables sensing \cite{Mehta-01}, adaptation \cite{Lan-01}, force generation \cite{Silva-01}, and structural organization \cite{Stuhrmann-01}.

Despite the importance of non-equilibrium dynamics at the microscale, there remain basic questions about the role of broken detailed balance in macroscopic systems composed of many interacting components. Do violations of detailed balance emerge at large scales? And if so, do such violations support higher-order biological functions, just as microscopic broken detailed balance drives molecular and cellular functions?

To answer these questions, we study large-scale patterns of activity in the brain. Notably, the human brain consumes up to 20\% of the body's energy in order to perform an array of cognitive functions, from computations and attention to planning and motor execution \cite{Harris-01, Lynn-05}, making it a promising system in which to probe for macroscopic broken detailed balance. Indeed, metabolic and enzymatic activity in the brain drives a number of non-equilibrium processes at the microscale, including neuronal firing \cite{Erecinska-01}, molecular cycles \cite{Norberg-01}, and cellular housekeeping \cite{Du-01}. One might therefore conclude that the brain -- indeed any living system -- must break detailed balance at large scales. However, by coarse-graining a system, one may average over non-equilibrium degrees of freedom, yielding ``effective" macroscopic dynamics that produce less entropy \cite{Esposito-01, Martinez-01} and regain detailed balance \cite{Egolf-01}. Thus, even though non-equilibrium processes are vital at molecular and cellular scales, it remains independently important to examine the role of broken detailed balance in the brain -- and in complex systems generally -- at large scales.

\section{Fluxes and broken detailed balance in the brain}

Here we develop tools to probe for and quantify broken detailed balance in macroscopic living systems. We apply our methods to analyze whole-brain dynamics from 590 healthy adults both at rest and across a suite of seven cognitive tasks, recorded using functional magnetic resonance imaging (fMRI) as part of the Human Connectome Project \cite{Van-01}. For each cognitive task (including rest), the time-series data consist of blood-oxygen-level-dependent (BOLD) fMRI signals from 100 cortical parcels \cite{Thomas-01}, which we concatenate across all subjects (see Appendix \ref{methods} for an extended description of the neural data).

We begin by visually examining whether the neural dynamics break detailed balance. To visualize the dynamics, we must project the time series onto two dimensions. For example, here we project the neural dynamics onto the first two principal components of the time-series data, which we compute after combining all data points across all subjects [Fig. \ref{fluxMap}(c)]. In fact, this projection defines a natural low-dimensional state space \cite{Cunningham-01}, capturing over 30\% of the variance in the neural activity (see Appendix \ref{projection}, Fig. \ref{PCA}). One can then probe for broken detailed balance by calculating the net fluxes of transitions between different regions of state space, a method proposed by Battle \textit{et al.} \cite{Battle-01} (see Appendix \ref{methods}). Moreover, we can repeat this analysis for different cognitive tasks to investigate whether the fluxes between neural states depend on the mental function being performed.

\begin{figure*}
\centering
\includegraphics[width = .8\textwidth]{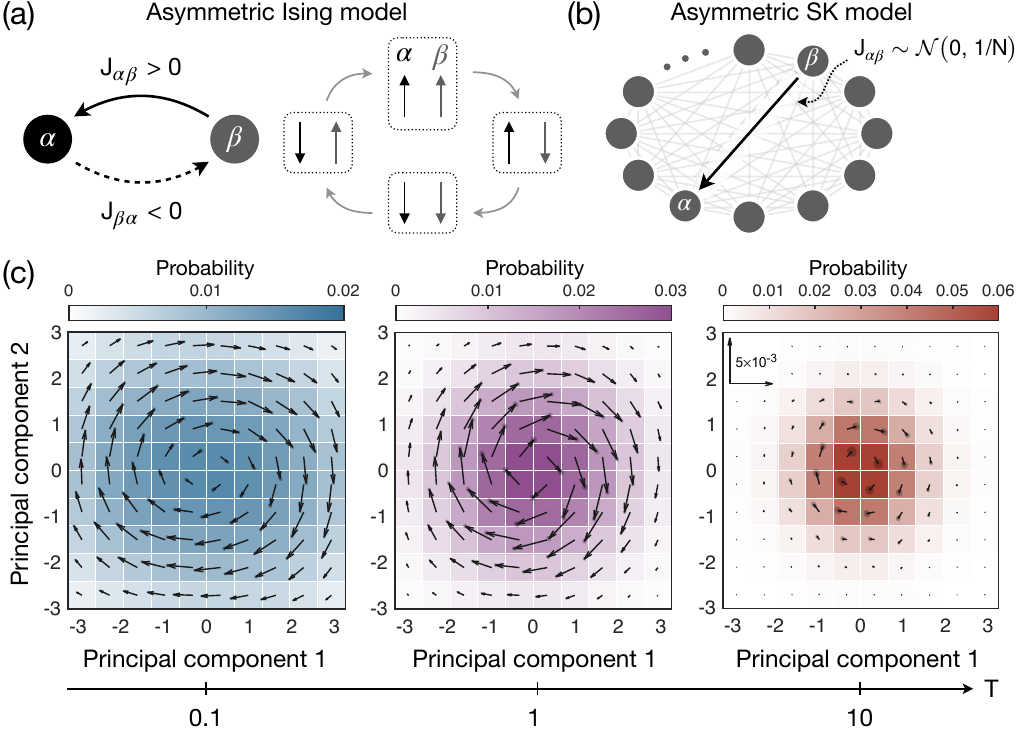} \\
\caption{\label{Ising} Emergence of macroscopic broken detailed balance in a complex system. (a) Two-spin Ising model with interactions $J_{\alpha\beta}$ representing the strength of the influence of spin $\beta$ on spin $\alpha$ (left). If the interactions are asymmetric (such that $J_{\alpha\beta} \neq J_{\beta\alpha}$), then the system exhibits a loop of flux between spin states (right). (b) Asymmetric Sherrington-Kirkpatrick (SK) model, wherein directed interactions are drawn independently from a zero-mean Gaussian with variance $1/N$, where $N$ is the size of the system. (c) For an asymmetric SK model with $N=100$ spins, we plot the probability distribution (color) and fluxes between states (arrows) for simulated time-series at temperatures $T = 0.1$ (left), $T = 1$ (middle), and $T = 10$ (right). In order to visualize the dynamics, the time series are projected onto the first two principal components of the combined data across all three temperatures. The scale is indicated in flux-per-time-step, and the disks represent two-standard-deviation confidence intervals that arise due to finite data (see Appendix \ref{methods}).}
\end{figure*}

We first consider the brain's behavior during resting scans, wherein subjects are instructed to remain still without executing a specific task. At rest, we find that the brain exhibits net fluxes between states [Fig. \ref{fluxMap}(d)], thereby establishing that neural dynamics break detailed balance at large scales. But are violations of detailed balance determined solely by the structural connections in the brain, or does the nature of broken detailed balance depend on the specific function being performed? 

To answer this question, we study task scans, wherein subjects respond to stimuli and commands that require attention, information processing, and physical and cognitive effort. For example, here we consider a gambling task in which subjects play a card guessing game for monetary reward. Interestingly, during the gambling task the fluxes between neural states are nearly an order of magnitude stronger than those present during rest [Fig. \ref{fluxMap}(e)]. Moreover, these fluxes combine to form a distinct loop in state-space, a characteristic feature of broken detailed balance in steady-state systems \cite{Zia-01}, and we verify that the brain does indeed operate at a stochastic steady state (see Appendix \ref{steady}, Fig. \ref{divergence}). To confirm that fluxes between neural states reflect broken detailed balance, and are not simply artifacts of our data processing, we show that if the time series are shuffled -- thereby destroying the temporal order of the system -- then the fluxes between states vanish and detailed balance is restored (see Appendix \ref{equilibrium}, Fig. \ref{shuffle}). Together, these results demonstrate that the brain fundamentally breaks detailed balance at large scales, and that the strength of broken detailed balance depends critically on the cognitive function being performed. \vspace{-10pt}

\section{Emergence of macroscopic broken detailed balance}

We have established that the brain breaks detailed balance at large scales, exhibiting net fluxes between macroscopic neural states. But can such large-scale violations of detailed balance emerge from fine-scale fluxes involving only one or two elements at a time? To answer these questions, we consider a canonical model of stochastic dynamics in complex systems. In the Ising model, the interactions between individual elements (or spins) are typically constrained to be symmetric, yielding simulated dynamics that obey detailed balance \cite{Newman-03}. However, connections in the brain -- from synapses between neurons to white matter tracts between entire brain regions -- are inherently asymmetric \cite{Kale-01}. If we allow for asymmetric interactions in the Ising model, then the system diverges from broken detailed balance at small scales, displaying loops of flux involving pairs of spins [Fig. \ref{Ising}(a)]. But can such fine-scale fluxes combine to generate large-scale violations of detailed balance?

To understand whether (and how) microscopic fluxes give rise to macroscopic broken detailed balance, we study a system of $N = 100$ spins (matching the 100 parcels in our neuroimaging data). Importantly, the system does not contain large-scale structure, with the interaction between each directed pair of spins drawn independently from a zero-mean Gaussian with variance $1/N$ [Fig. \ref{Ising}(b)]. This model is the asymmetric generalization of the Sherrington-Kirkpatrick (SK) model of a spin glass \cite{Sherrington-01}. After simulating the system at three different temperatures, we perform the same procedure that we applied to the neuroimaging data (Fig. \ref{fluxMap}): projecting the time-series onto the first two principal components of the combined data and measuring net fluxes in this low-dimensional state space.

At high temperature, stochastic fluctuations dominate the system, and we only observe weak fluxes between states [Fig. \ref{Ising}(c), right]. By contrast, as the temperature decreases, the interactions between spins overcome the stochastic fluctuations, giving rise to clear loops of flux [Fig. \ref{Ising}(c), middle and left]. These loops of flux demonstrate that asymmetries in the fine-scale interactions between elements alone can give rise to large-scale broken detailed balance. Moreover, by varying the strength of microscopic interactions, a single system can transition from exhibiting small violations of detailed balance to dramatic loops of flux, just as observed in the brain during different cognitive tasks [Fig. \ref{fluxMap}(d-e)].

\section{Quantifying broken detailed balance: Entropy production}

While fluxes in state space reveal violations of detailed balance, quantifying this behavior requires measuring the ``distance" of a system from detailed balance. One such measure is entropy production, the central concept of non-equilibrium statistical mechanics \cite{Seifert-01, Esposito-01, Roldan-01, Gnesotto-01}. In microscopic systems, the rate at which entropy is produced -- that is, the distance of the system from detailed balance -- can often be directly related to the consumption of energy needed to drive cellular and sub-cellular functions \cite{Gnesotto-01, Mehta-01, Lan-01}. In macroscopic systems, this physical entropy production is lower-bounded by an information-theoretic notion of entropy production, which can be estimated simply by observing a system's coarse-grained dynamics \cite{Roldan-01, Esposito-01}. For example, consider a system with joint transition probabilities $P_{ij} = \text{Prob}[x_{t-1} = i,\, x_t = j]$, where $x_t$ is the state of the system at time $t$. If the dynamics are Markovian (as, for instance, is true for the Ising system), then the information entropy production is given by
\begin{equation}
\label{S}
\dot{S} = \sum_{ij} P_{ij} \log \frac{P_{ij}}{P_{ji}},
\end{equation}
where the sum runs over all states $i$ and $j$. For simplicity, we refer to the information entropy production above simply as entropy production, not to be confused with the physical production of entropy at the microscale.

\begin{figure*}
\centering
\includegraphics[width = .8\textwidth]{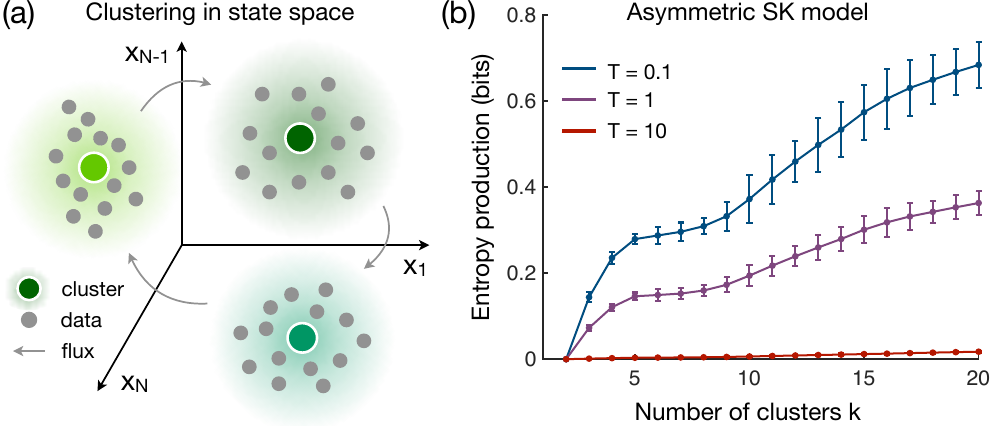} \\
\caption{\label{clustering} Estimating entropy production using hierarchical clustering. (a) Schematic of clustering procedure, where axes represent the activities of individual components (e.g., brain regions in the neuroimaging data or spins in the Ising model), points reflect individual states observed in the time-series, shaded regions define clusters (or coarse-grained states), and arrows illustrate possible fluxes between clusters. (b) Entropy production in the asymmetric SK model as a function of the number of clusters $k$ for the same time-series studied in Fig. \ref{Ising}(c), with error bars reflecting two-standard-deviation confidence intervals that arise due to finite data (see Appendix \ref{methods}).}
\end{figure*}

Inspecting Eq. (\ref{S}), it becomes clear why entropy production is a natural measure of broken detailed balance: It is the Kullback-Leibler divergence between the forward transition probabilities $P_{ij}$ and the reverse transition probabilities $P_{ji}$ \cite{Cover-01}. If the system obeys detailed balance (that is, if $P_{ij} = P_{ji}$ for all pairs of states $i$ and $j$), then the entropy production vanishes. Conversely, any violation of detailed balance (that is, any flux of transitions such that $P_{ij} \neq P_{ji}$) leads to an increase in entropy production.

Calculating the entropy production requires estimating the transition probabilities $P_{ij}$. However, for complex systems the number of states grows exponentially with the size of the system, making a direct estimate of the entropy production infeasible. To overcome this hurdle, we employ a hierarchical clustering algorithm that groups similar states in a time series into a single cluster, yielding a reduced number of coarse-grained states [Fig. \ref{clustering}(a); see Appendix \ref{methods}]. By choosing these clusters hierarchically \cite{Lamrous-01}, we prove that the estimated entropy production can only increase with the number of coarse-grained states -- that is, as our description of the system becomes more fine-grained (ignoring finite data effects; see Appendix \ref{hierarchical}). Indeed, across all temperatures in the Ising system, we verify that the estimated entropy production increases with the number of clusters $k$ [Fig. \ref{clustering}(b)]. Furthermore, as the temperature decreases, the entropy production increases [Fig. \ref{clustering}(b)], thereby capturing the growing violations of detailed balance at low versus high temperatures [Fig. \ref{Ising}(c)].

\begin{figure*}
\centering
\includegraphics[width = .65\textwidth]{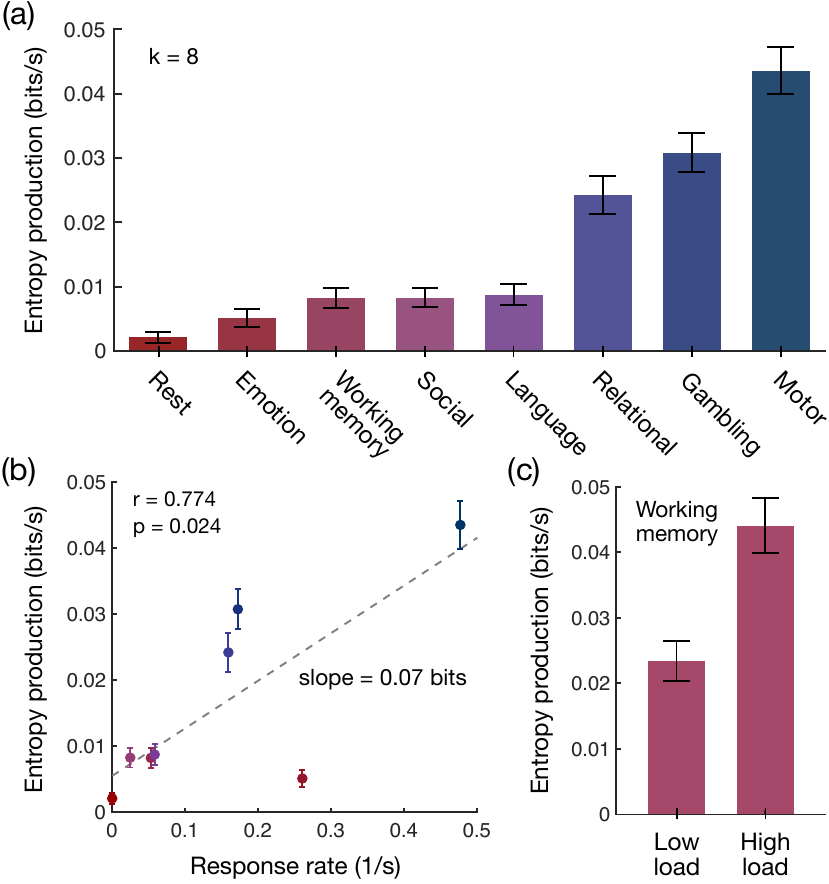} \\
\caption{\label{entProd} Entropy production in the brain varies with physical and cognitive demands. (a) Entropy production at rest and during seven cognitive tasks, estimated using hierarchical clustering with k = 8 clusters. (b) Entropy production as a function of response rate (i.e., the frequency with which subjects are asked to physically respond) for the tasks listed in panel (a). Each response induces an average $0.07\pm 0.03$ bits of produced entropy (Pearson correlation $r = 0.774$, $p = 0.024$). (c) Entropy production for low cognitive load and high cognitive load conditions in the working memory task, where low and high loads represent 0-back and 2-back conditions, respectively, in an n-back task. The brain produces significantly more entropy during high-load than low-load conditions (one-sided \textit{t}-test, $p < 0.001$, $t > 10$, $df = 198$). Across all panels, raw entropy productions [Eq. (\ref{S})] are divided by the fMRI repetition time $\Delta t = 0.72$ s to compute an entropy production rate, and error bars reflect two-standard-deviation confidence intervals that arise due to finite data (see Appendix \ref{methods}).}
\end{figure*}

\section{Entropy production in the human brain}

We are now prepared to investigate the extent to which the brain breaks detailed balance when performing different functions. We study seven tasks, each of which engages a specific cognitive process and associated anatomical system: emotional processing, working memory, social inference, language processing, relational matching, gambling, and motor execution \cite{Barch-01}. To estimate the entropy production of the neural dynamics, we cluster the neuroimaging data (combined across all subjects and task settings, including rest) into $k=8$ coarse-grained states, the largest number for which all transitions were observed at least once in each task (see Appendix \ref{choosing_k}, Fig. \ref{num_states}). Across all tasks and rest, we find that the neural dynamics produce entropy, confirming that the brain breaks detailed balance at large scales [Fig. \ref{entProd}(a)]. Specifically, for all settings the entropy production is significantly greater than the noise floor that arises due to finite data (one-sided \textit{t}-test with $p < 0.001$).

Interestingly, the neural dynamics produce more entropy during all of the cognitive tasks than at rest [Fig. \ref{entProd}(a)]. In the motor task, for example -- wherein subjects are prompted to perform specific physical movements -- the entropy production is 20 times larger than for resting-state dynamics [Fig. \ref{entProd}(a)]. In fact, while each cognitive task induces a unique pattern of fluxes between neural states (see Appendix \ref{flux_nets}, Fig. \ref{fluxNets}), these fluxes nearly vanish during resting scans [Fig. \ref{fluxNets}(b)]. Thus, we find that the extent to which the brain breaks detailed balance -- and the manner in which it does so -- depend critically on the specific task being performed.

The above results demonstrate that the brain breaks detailed balance at large scales as it executes physical movements, processes information, and performs cognitive functions. Indeed, just as energy is expended at the microscale to break detailed balance \cite{Gnesotto-01}, one might expect violations of detailed balance in neural dynamics to increase with physical and cognitive exertion. To test the first hypothesis -- that broken detailed balance in the brain is associated with physical effort -- we compare the brain's entropy production in each task with the frequency of physical movements [Fig. \ref{entProd}(b)]. Across tasks, we find that entropy production does in fact increase with the frequency of motor responses, with each response yielding an additional $0.07\pm 0.03$ bits of information entropy. Additionally, we confirm that this relationship between entropy production and physical effort also holds at the level of individual humans (see Appendix \ref{individual}, Fig. \ref{entProd_ind}).

To study the impact of cognitive effort and information processing on broken detailed balance, we focus on the working memory task, which splits naturally into two conditions with high and low cognitive loads. Importantly, the frequency of physical responses is identical across the two conditions, thereby controlling for the effect of physical effort studied previously. We find that the brain operates further from detailed balance when exerting more cognitive effort [Fig. \ref{entProd}(c)], with the high-load condition inducing a two-fold increase in entropy production over the low-load condition. Moreover, at the level of individuals, we find that entropy production increases with task errors (see Appendix \ref{individual}), once again indicating that violations of detailed balance intensify with cognitive demand.

Finally, we verify that our results do not depend on the Markov assumption in Eq. (\ref{S}) (see Appendix \ref{assumption}, Fig. \ref{Markov}), are robust to reasonable variation in the number of clusters $k$ (see Appendix \ref{vary_k}, Fig. \ref{robust_k}), and cannot be explained by head motion in the scanner (a common confound in fMRI studies \cite{Friston-05}) nor variance in the activity time-series (see Appendix \ref{robust}, Fig. \ref{DVARS}). Together, these findings demonstrate that large-scale violations of detailed balance in the brain robustly increase with measures of both physical effort and cognitive demand. These conclusions, in turn, suggest that broken detailed balance in macroscopic systems may support higher-order biological functions. \vspace{-10pt}

\section{Conclusions}

In this study, we describe a method for investigating macroscopic broken detailed balance by quantifying entropy production in living systems. While microscopic non-equilibrium processes are known to be vital for molecular and cellular operations \cite{Brangwynne-01, Yin-01, Huang-01, Mehta-01, Lan-01, Silva-01, Stuhrmann-01}, here we show that broken detailed balance also arises at large scales in complex living systems. Analyzing whole-brain imaging data, we demonstrate that the human brain breaks detailed balance at large scales. Moreover, we find that the brain's entropy production (that is, its distance from detailed balance) varies critically with the specific function being performed, increasing with both physical and cognitive demands.

These results open the door for a number of important future directions. For example, given that large-scale violations of detailed balance can emerge from fine-scale asymmetries in a system (Fig. \ref{Ising}), future research should examine the relationship between broken detailed balance in the brain and asymmetries in the structural connectivity between brain regions. Similarly, given the intimate relationship between broken detailed balance and energy consumption at the molecular and cellular scales, it is natural to investigate whether violations of detailed balance at large scales are associated with fluctuations in neural metabolism.

More generally, we remark that the presented framework is non-invasive, applying to any system with time-series data. Furthermore, these methods can be used to study stochastic steady-state dynamics, rather than deterministic dynamics that trivially break detailed balance. Thus, the framework does not only apply to the brain, but can be used broadly to investigate broken detailed balance in other complex systems, including emergent behavior in human and animal populations \cite{Castellano-01}, correlated patterns of neuronal firing \cite{Palva-01}, and collective activity in molecular and cellular networks \cite{Koenderink-01, Van-02}.

\begin{acknowledgments}
The authors thank Erin Teich, Pragya Srivastava, Jason Kim, and Zhixin Lu for feedback on earlier versions of this manuscript. The authors acknowledge support from the John D. and Catherine T. MacArthur Foundation, the ISI Foundation, the Paul G. Allen Family Foundation, the Army Research Laboratory (W911NF-10-2-0022), the Army Research Office (Bassett-W911NF-14-1-0679, Falk-W911NF-18-1-0244, Grafton-W911NF-16-1-0474, DCIST- W911NF-17-2-0181), the Office of Naval Research, the National Institute of Mental Health (2-R01-DC-009209-11, R01-MH112847, R01-MH107235, R21-M MH-106799), the National Institute of Child Health and Human Development (1R01HD086888-01), National Institute of Neurological Disorders and Stroke (R01 NS099348), and the National Science Foundation (NSF PHY-1554488, BCS-1631550, and NCS-FO-1926829).
\end{acknowledgments}

\appendix

\section*{CITATION DIVERSITY STATEMENT}

Recent work in neuroscience and other fields has identified a bias in citation practices such that papers from women and other minorities are under-cited relative to the number of such papers in the field \cite{Dworkin-01, Caplar-01}. Here we sought to proactively consider choosing references that reflect the diversity of the field in thought, form of contribution, gender, race, geographic location, and other factors. Excluding self-citations to the authors of this paper and single-author citations, the first and last authors of references are 58\% male/male, 21\% female/male, 14\% male/female, and 7\% female/female. We look forward to future work that could help us better understand how to support equitable practices in science.

\section{Methods}
\label{methods}

\subsection{Calculating fluxes}

Consider time-series data gathered in a time window $t_{\text{tot}}$, and let $n_{ij}$ denote the number of observed transitions from state $i$ to state $j$. The flux rate from state $i$ to state $j$ is given by $\omega_{ij} = (n_{ij} - n_{ji})/t_{\text{tot}}$. For the flux currents in Figs. \ref{fluxMap}(d-e) and \ref{Ising}(c), the states of the system are points $(x, y)$ in two-dimensional space, and the state probabilities are estimated by $p(x,y) = t_{(x,y)}/t_{\text{tot}}$, where $t_{(x,y)}$ is the time spent in state $(x,y)$. The magnitude and direction of the flux through a given state $(x,y)$ is defined by the flux vector \cite{Battle-01}
\begin{equation}
\bm{u}(x, y) = \frac{1}{2}\Bigg(\begin{array}{c} \omega_{(x-1,y), (x,y)} + \omega_{(x,y), (x+1,y)} \\ \omega_{(x,y-1), (x,y)} + \omega_{(x,y), (x,y+1)} \end{array} \Bigg).
\end{equation}
In a small number of cases, two consecutive states in the observed time-series $\bm{x}(t) = (x(t), y(t))$ and $\bm{x}(t+1) = (x(t+1), y(t+1))$ are not adjacent in state space. In these cases, we perform a linear interpolation between $\bm{x}(t)$ and $\bm{x}(t+1)$ in order to calculate the fluxes between adjacent states.

\subsection{Estimating finite-data errors using trajectory bootstrapping}

The finite length of time-series data limits the accuracy with which quantities -- such as entropy production and the fluxes between states -- can be estimated. In order to calculate error bars on all estimated quantities, we apply trajectory bootstrapping \cite{Shannon-01, Battle-01}. We first record the list of transitions
\begin{equation}
I = \left(\begin{array}{cc} i_1 & i_2 \\
i_2 & i_3 \\
\vdots & \vdots \\
i_{L-1} & i_L \end{array} \right),
\end{equation}
where $i_{\ell}$ is the ${\ell}^{\text{th}}$ state in the time-series, and $L$ is the length of the time-series. From the transition list $I$, one can calculate all of the desired quantities; for instance, the fluxes are estimated by
\begin{equation}
\omega_{ij} = \frac{1}{t_{\text{tot}}} \sum_{\ell} \delta_{i,I_{\ell,1}}\delta_{j,I_{\ell,2}} - \delta_{j,I_{\ell,1}}\delta_{i,I_{\ell,2}}.
\end{equation}
We remark that when analyzing the neural data, although we concatenate the time-series across subjects, we only include transitions in $I$ that occur within the same subject. That is, we do not include the transitions between adjacent subjects in the concatenated time-series.

\begin{figure}
\centering
\includegraphics[width = .35\textwidth]{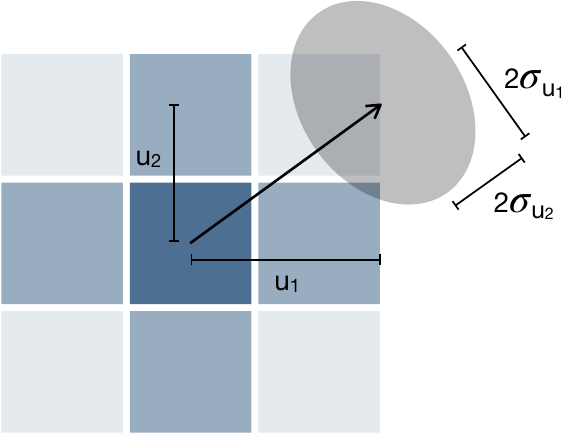} \\
\caption{\label{fluxVec} Visualizing flux vectors. Schematic demonstrating how we illustrate the flux of transitions through a state (vector) and the errors in estimating the flux due to finite data (ellipse).}
\end{figure}

To calculate errors, we construct bootstrap trajectories (of the same length $L$ as the original data) by sampling the rows in $I$ with replacement. For example, by calculating the entropy production in each of the bootstrap trajectories, we are able estimate the size of finite-data errors in Figs. \ref{clustering}(b) and \ref{entProd}. Similarly, to compute errors for the flux vectors $\bm{u}(\bm{x})$ in Figs. \ref{fluxMap}(d-e) and \ref{Ising}(c), we first estimate the covariance matrix $\text{Cov}(u_1(\bm{x}), u_2(\bm{x}))$ by averaging over bootstrapped trajectories. Then, for each flux vector, we visualize its error by plotting an ellipse with axes aligned with the eigenvectors of the covariance matrix and radii equal to twice the square root of the corresponding eigenvalues (Fig. \ref{fluxVec}). All errors throughout the manuscript are calculated using 100 bootstrap trajectories.

The finite data length also induces a noise floor for each quantity, which is present even if the temporal order of the time-series is destroyed. To estimate the noise floor, we construct bootstrap trajectories by sampling individual data points from the time-series. We contrast these bootstrap trajectories with those used to estimate errors above, which preserve transitions by sampling the rows in $I$. The noise floor, which is calculated for each quantity by averaging over the bootstrap trajectories, is then compared with the estimated quantities. For example, rather than demonstrating that the average entropy productions in Fig. \ref{entProd}(a) are greater than zero, we establish that the distribution over entropy productions is significantly greater than the noise floor using a one-sided \textit{t}-test with $p < 0.001$.

\subsection{Simulating the asymmetric Ising model}

The asymmetric Ising model is defined by a (possibly asymmetric) interaction matrix $J$, where $J_{\alpha \beta}$ represents the influence of spin $\beta$ on spin $\alpha$ [Fig. \ref{Ising}(a)], and a temperature $T \ge 0$ that tunes the strength of stochastic fluctuations. Here, we study a system with $N=100$ spins, where each directed interaction $J_{\alpha\beta}$ is drawn independently from a zero-mean Gaussian with variance $1/N = 0.01$ [Fig. \ref{Ising}(b)]. One can additionally include external fields $h_{\alpha}$, but for simplicity here we set them to zero. The state of the system is defined by a vector $\bm{x} = (x_1,\hdots, x_N)$, where $x_{\alpha} = \pm 1$ is the state of spin $\alpha$. To generate time series, we employ Glauber dynamics with synchronous updates, a common Monte Carlo method for simulating Ising systems \cite{Newman-03}. Specifically, given the state of the system $\bm{x}(t)$ at time $t$, the probability of spin $\alpha$ being ``up" at time $t+1$ (that is, the probability that $x_\alpha(t+1) = 1$) is given by
\begin{widetext}
\begin{equation}
\label{Glauber}
\text{Prob}[x_{\alpha}(t+1) = 1\,|\, \bm{x}(t)] = \frac{\exp\left(\frac{1}{T}\sum_{\beta} J_{\alpha\beta} x_{\beta}(t)\right)}{\exp\left(\frac{1}{T}\sum_{\beta} J_{\alpha\beta} x_{\beta}(t)\right) + \exp\left(-\frac{1}{T}\sum_{\beta} J_{\alpha\beta} x_{\beta}(t)\right)}.
\end{equation}
\end{widetext}
Stochastically updating each spin $\alpha$ according to Eq. (\ref{Glauber}), one arrives at the new state $\bm{x}(t+1)$. For each temperature in the Ising calculations in Figs. \ref{Ising}(c) and \ref{clustering}(b), we generate a different time-series of length $L = 100,000$ with $10,000$ trials of burn-in.

\subsection{Hierarchical clustering}

To estimate the entropy production of a system, one must first calculate the transition probabilities $P_{ij} = n_{ij}/(L-1)$. For complex systems, the number of states $i$ (and therefore the number of transitions $i\rightarrow j$) grows exponentially with the size of the system $N$. For example, in the Ising model each spin $\alpha$ can take one of two values ($x_{\alpha} = \pm 1$), leading to $2^N$ possible states and $2^{2N}$ possible transitions. In order to estimate the transition probabilities $P_{ij}$, one must observe each transition $i\rightarrow j$ at least once, which requires significantly reducing the number of states in the system. Rather than defining coarse-grained states \textit{a priori}, complex systems (and the brain in particular) often admit natural coarse-grained descriptions that are uncovered through dimensionality-reduction techniques \cite{Cunningham-01, Cornblath-01, Liu-03}.

Although one can use any coarse-graining technique to implement our framework and estimate entropy production, here we employ hierarchical $k$-means clustering for two reasons: (i) generally, $k$-means is perhaps the most common and simplest clustering algorithm, with demonstrated effectiveness fitting neural dynamics \cite{Cornblath-01, Liu-03}; and (ii) specifically, by defining the clusters hierarchically we prove that the estimated entropy production becomes more accurate as the number of clusters increases (ignoring finite data effects; Fig. \ref{coarse}).

In $k$-means clustering, one begins with a set of states (for example, those observed in our time-series) and a number of clusters $k$. Each observed state $\bm{x}$ is randomly assigned to a cluster $i$, and one computes the centroid of each cluster. On the following iteration, each state is re-assigned to the cluster with the closest centroid (here we use cosine similarity to determine distance). This process is repeated until the cluster assignments no longer change. In a hierarchical implementation, one begins with two clusters; then one cluster is selected (typically the one with the largest spread in its constituent states) to be split into two new clusters, thereby defining a total of three clusters. This iterative splitting is continued until one reaches the desired number of clusters $k$. In Appendix \ref{hierarchical}, we show that hierarchical clustering provides an increasing lower-bound on the entropy production; and in Appendix \ref{choosing_k}, we demonstrate how to choose the number of clusters $k$.

\subsection{Neural data}

The whole-brain dynamics used in this study are measured and recorded using blood-oxygen-level-dependent (BOLD) functional magnetic resonance imaging (fMRI) collected from 590 healthy adults as part of the Human Connectome Project \cite{Van-01, Barch-01}. For each subject, recordings were taken during seven different cognitive tasks and also during rest. BOLD fMRI estimates neural activity by calculating contrasts in blood oxygen levels, without relying on invasive injections and radiation \cite{Raichle-01}. Specifically, blood oxygen levels (reflecting neural activity) are measured within three-dimensional non-overlapping voxels, spatially contiguous collections of which each represent a distinct brain region (or parcel). Here, we consider a parcellation that divides the cortex into 100 brain regions that are chosen to optimally capture the functional organization of the brain \cite{Thomas-01}. After processing the signal to correct for sources of systematic noise such as head motion (see Appendix \ref{processing}), the activity of each brain region is discretized in time, yielding a time-series of neural activity. For each subject, the shortest scan (corresponding to the emotional processing task) consists of 176 discrete measurements in time. In order to control for variability in data size across tasks, for each subject we only study the first 176 measurements in each task.

\begin{figure}
\centering
\includegraphics[width = .4\textwidth]{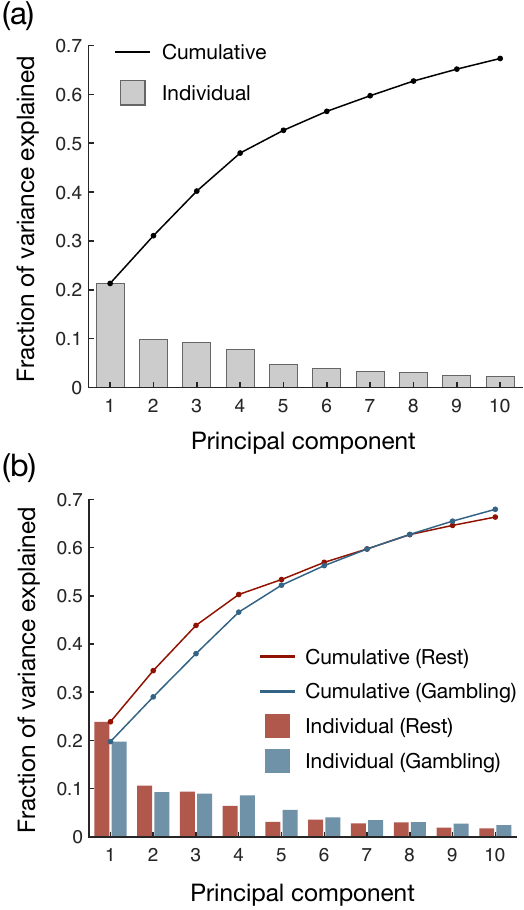} \\
\caption{\label{PCA} PCA reveals low-dimensional embedding of neural dynamics. (a) Cumulative fraction of variance explained by first ten principal components (line) and explained variance for each individual principal component (bars) in the combined rest and gambling data. (b) For the same principal components (calculated for the combined rest and gambling data), we plot the cumulative fraction of variance explained (lines) and individual explained variance (bars) for the rest (red) and gambling (blue) data.}
\end{figure}

\section{Low-dimensional embedding using PCA}
\label{projection}

In order to visualize net fluxes between states in a complex system, we must project the dynamics onto two dimensions. While any pair of dimensions can be used to probe for broken detailed balance, a natural choice is the first two principal components of the time-series data. Indeed, principal component analysis has been widely used to uncover low-dimensional embeddings of large-scale neural dynamics \cite{Cunningham-01, Song-01}. Combining the time-series data from the rest and gambling task scans (that is, the data studied in Fig. \ref{fluxMap}), we find that the first two principal components capture over 30\% of the total variance in the observed recordings [Fig. \ref{PCA}(a)], thereby comprising a natural choice for two-dimensional projections. Moreover, we confirm that the projected dynamics capture approximately the same amount of variance in both the rest and gambling tasks, confirming that PCA is not overfitting the neural dynamics in one task or another [Fig. \ref{PCA}(b)].

\section{The brain operates at a stochastic steady state}
\label{steady}

\begin{figure}[t!]
\centering
\includegraphics[width = .4\textwidth]{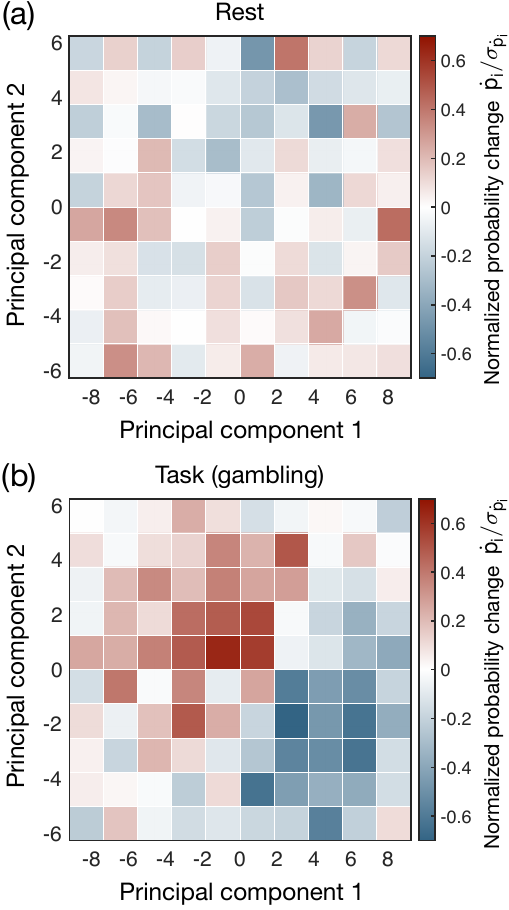} \\
\caption{\label{divergence} Small changes in state probabilities imply steady-state dynamics. Change in state probabilities $\dot{p}_i$, normalized by the standard deviation $\sigma_{\dot{p}_i}$, plotted as a function of the first two principal components at rest (a) and during the gambling task (b).}
\end{figure}

Some of the tools and intuitions developed in traditional statistical mechanics to study equilibrium systems have recently been generalized to systems that operate at non-equilibrium steady states \cite{Seifert-02}. For example, Evans \textit{et al.} generalized the second law of thermodynamics to non-equilibrium steady-state systems by discovering the (steady state) fluctuation theorem \cite{Evans-02}. More recently, Dieterich \textit{et al.} showed that, by mapping their dynamics to an equilibrium system at an effective temperature, some non-equilibrium steady-state systems are governed by a generalization of the fluctuation-dissipation theorem \cite{Dieterich-01}. Thus, it is both interesting and practical to investigate whether the brain operates at a non-equilibrium steady state. We remark that by ``non-equilibrium" we refer to the breaking of detailed balance at large scales, not the obvious non-equilibrium nature of the brain at the cellular and molecular scales.

We establish in Figs. \ref{fluxMap} and \ref{entProd} that the brain breaks detailed balance. To determine if the brain functions at a steady state, we must examine whether its state probabilities are stationary in time; that is, letting $p_i$ denote the probability of state $i$, we must determine whether $\dot{p}_i = dp_i /dt = 0$ for all states $i$. The change in the probability of a state is equal to the net rate at which transitions flow into versus out of a state. For the two-dimensional dynamics studied in Fig. \ref{fluxMap}, this relation takes the form
\begin{align}
\frac{dp_{(x,y)}}{dt} = \,\, &\omega_{(x-1,y),(x,y)} - \omega_{(x,y),(x+1,y)} \nonumber \\
& + \omega_{(x,y-1),(x,y)} - \omega_{(x,y),(x,y+1)},
\end{align}
where $\omega_{ij} = (n_{ij} - n_{ji})/t_{\text{tot}}$ is the flux rate from state $i$ to state $j$, $n_{ij}$ is the number of observed transitions $i\rightarrow j$, and $t_{\text{tot}}$ is the temporal duration of the time-series \cite{Battle-01}.

Here, we calculate the changes in state probabilities for both the rest and gambling scans. Across all states in both task conditions, we find that these changes are indistinguishable from zero when compared to statistical noise (Fig. \ref{divergence}). Specifically, the changes in state probabilities are much less than twice their standard deviations, indicating that they cannot be significantly distinguished from zero with a $p$-value less than $0.05$. Combined with the results from Figs. \ref{fluxMap} and \ref{entProd}, the stationarity of the neural state probabilities demonstrates that the brain operates at a non-equilibrium steady-state.

\section{Shuffling time-series restores detailed balance}
\label{equilibrium}

\begin{figure}
\centering
\includegraphics[width = .4\textwidth]{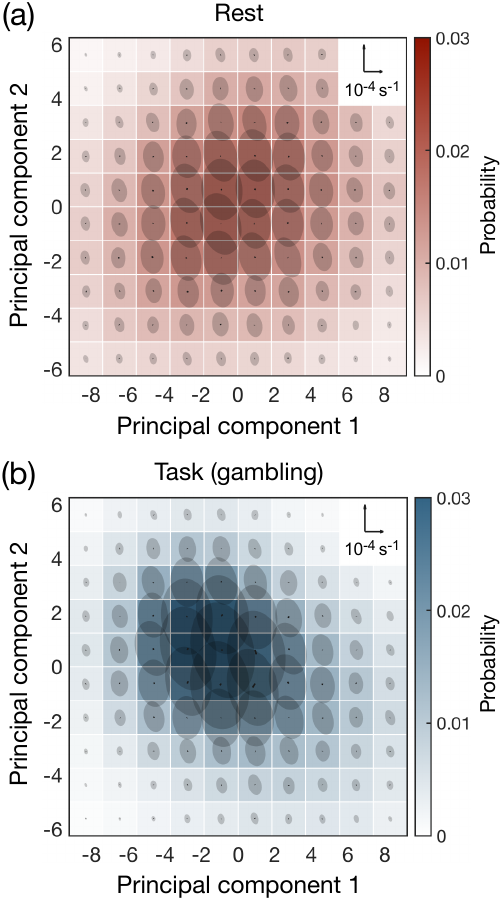} \\
\caption{\label{shuffle} Shuffled data do not exhibit fluxes between neural states. Probability distribution (color) and nearly imperceivable fluxes between states (arrows) for neural dynamics, which are shuffled and projected onto the first two principal components, both at rest (a) and during a gambling task (b). The flux scale is indicated in the upper right, and the disks represent two-standard-deviation confidence intervals that arise due to finite data (see Appendix \ref{methods}).}
\end{figure}

\begin{figure*}
\centering
\includegraphics[width = .7\textwidth]{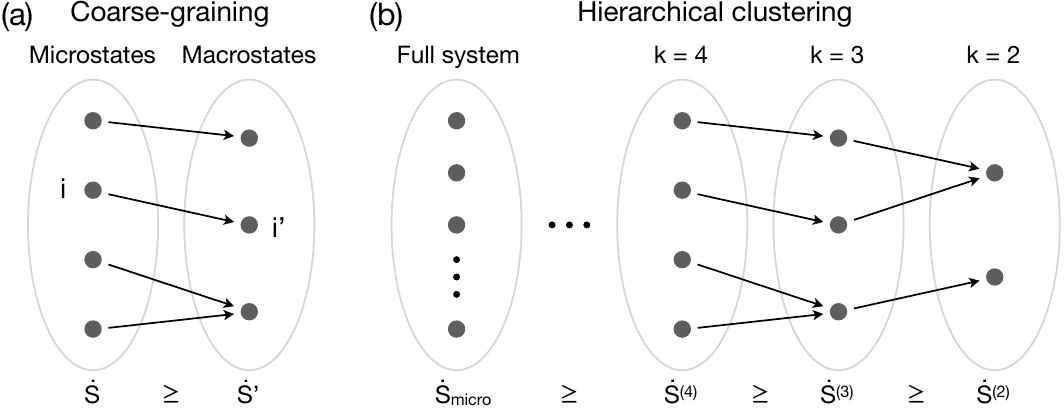} \\
\caption{\label{coarse} Hierarchy of lower bounds on entropy production. (a) Coarse-graining is defined by a surjective map from a set of microstates $\{i\}$ to a set of macrostates $\{i'\}$. Under coarse-graining the entropy production cannot increase. (b) In hierarchical clustering, states are iteratively combined to form new coarse-grained states (or clusters). Each iteration defines a coarse-graining from $k$ states to $k-1$ states, thereby forming a hierarchy of lower bounds on the entropy production.}
\end{figure*}

In Fig. \ref{fluxMap}, we demonstrate that the brain breaks detailed balance by exhibiting net fluxes between states. Here we demonstrate that if the temporal order of the neural data is destroyed (say, by shuffling the time-series), then the fluxes vanish and detailed balance is restored. Specifically, for both the rest and gambling task scans, we generate 100 surrogate time-series with the order of the data randomly shuffled. Averaging across these shuffled time-series, we find that the fluxes between states are vanishingly small compared to statistical noise (Fig. \ref{shuffle}), thus illustrating that the system has returned to detailed balance. We remark that other common surrogate data techniques, such as the random phases and amplitude adjusted Fourier transform surrogates, are not applicable here because they preserve the temporal structure of the time-series data \cite{Lancaster-01}.

\section{Estimating entropy production using hierarchical clustering}
\label{hierarchical}

Complex systems are often high-dimensional, with the number of possible states or configurations growing exponentially with the size of the system. In order to estimate the information entropy production [Eq. (\ref{S})] of a complex system, we must reduce the number of states through the use of coarse-graining, or dimensionality reduction, techniques. Interestingly, the entropy production admits a number of strong properties under coarse-graining \cite{Esposito-01, Martinez-01, Seifert-01, Roldan-01}. Of particular interest is the fact that the entropy production can only decrease under coarse-graining \cite{Esposito-01}. Specifically, given two descriptions of a system, a ``microscopic" description with states $\{i\}$ and a ``macroscopic" description with states $\{i'\}$, we say that the second description is a coarse-graining of the first if there exists a surjective map from the microstates $\{i\}$ to the macrosctates $\{i'\}$ [that is, if each microstate $i$ maps to a unique macrostate $i'$; Fig. \ref{coarse}(a)]. Given such a coarse-graining, Esposito showed \cite{Esposito-01} that the entropy production of the macroscopic description $\dot{S}'$ can be no larger than that of the microscopic description $\dot{S}$; in other words, the coarse-grained entropy production provides a lower bound for the original value, such that $\dot{S}' \le \dot{S}$.

The monotonic decrease of the entropy production under coarse-graining implies two desirable mathematical results. First, if one finds that any coarse-grained description of a system breaks detailed balance (that is, if the entropy production at any level of coarse-graining is significantly greater than zero), then one has immediately established that the full microscopic system breaks detailed balance. Thus, even without knowledge of the microscopic non-equilibrium processes at play, one can establish that the brain fundamentally breaks detailed balance at small scales simply by observing violations of detailed balance at large scales (Figs. \ref{fluxMap} and \ref{entProd}).

By extending this logic, here we show that hierarchical clustering provides systematic improvements to the entropy production estimates. In hierarchical clustering, each cluster (or coarse-grained state) at one level of description (with $k$ clusters) maps to a unique cluster at the level below [with $k-1$ clusters;  Fig. \ref{coarse}(b)]. This process can either be carried out by starting with a large number of clusters and then iteratively picking pairs of clusters to combine (known as agglomerative clustering), or by starting with a small number of clusters and then iteratively picking one cluster to split into two (known as divisive clustering, which we employ in our analysis) \cite{Cohen-03}. In both cases, the mapping from $k$ clusters to $k-1$ clusters is surjective, thereby representing a coarse-graining of the system, as defined previously. Thus, letting $\dot{S}^{(k)}$ denote the entropy production estimated with $k$ clusters, hierarchical clustering defines a hierarchy of lower bounds on the microscopic entropy production $\dot{S}_{\text{micro}}$:
\begin{equation}
0 = \dot{S}^{(1)} \le \dot{S}^{(2)} \le \dot{S}^{(3)} \le \hdots \le \dot{S}_{\text{micro}}.
\end{equation}
This hierarchy, in turn, demonstrates that the estimated entropy production $\dot{S}^{(k)}$ becomes larger (and thus more accurate) with increasing $k$.

\begin{figure*}
\centering
\includegraphics[width = \textwidth]{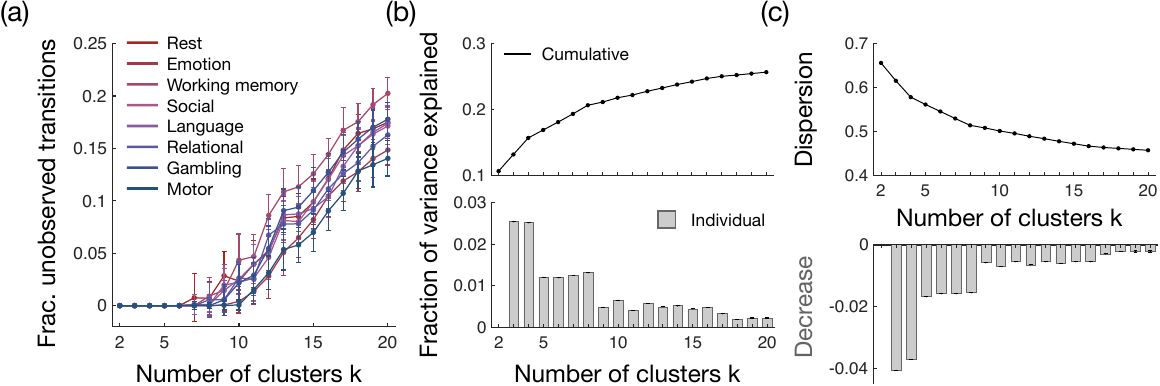} \\
\caption{\label{num_states} Choosing the number of coarse-grained states $k$. (a) Fraction of the $k^2$ state transitions that remain unobserved after hierarchical clustering with $k$ clusters for the different tasks. Error bars represent two-standard-deviation confidence intervals that arise due to finite data. (b) Percent variance explained (top) and the increase in explained variance from $k-1$ to $k$ clusters (bottom) as functions of $k$. (c) Dispersion, or the average distance between data points within a cluster (top), and the decrease in dispersion from $k-1$ to $k$ clusters (bottom) as functions of $k$.}
\end{figure*}

We remark that the discussion above neglects finite data effects. We recall that estimating the entropy production requires first estimating the transition probabilities $P_{ij}$ from state $i$ to state $j$. This means that for Markov systems with $k$ clusters, one must estimate $k^2$ different probabilities. Thus, while increasing $k$ improves the accuracy of the estimated entropy production in theory, in practice increasing $k$ eventually leads to sampling issues that decrease the accuracy of the estimate. Given these competing influences, when analyzing real data the goal should be to choose $k$ such that it is as large as possible while still providing accurate estimates of the transition probabilities. We discuss a systematic strategy for choosing $k$ in the following section.

\section{Choosing the number of coarse-grained states}
\label{choosing_k}

As discussed above, when calculating the entropy production, we wish to choose a number of coarse-grained states $k$ that is as large as possible while still arriving at an accurate estimate of the transition probabilities. One simple condition for estimating each transition probability $P_{ij}$ is that we observe the transition $i\rightarrow j$ at least once in the time-series. For all of the different tasks, Fig. \ref{num_states}(a) shows the fraction of the $k^2$ state transitions that are left unobserved after coarse-graining with $k$ clusters. We find that $k = 8$ is the largest number of clusters for which the fraction of unobserved transitions equals zero (within statistical errors) for all tasks; that is, the largest number of clusters for which all state transitions across all tasks were observed at least once. For this reason, we use $k=8$ coarse-grained states to analyze the brain's entropy production (Fig. \ref{entProd}).

Interestingly, we find that $k=8$ coarse-grained states is a good choice for two additional reasons. The first comes from studying the amount of variance explained by $k$ clusters [Fig. \ref{num_states}(b)]. We find that the increase in explained variance from $k-1$ to $k$ clusters is roughly constant for $k = 3$ and $4$, then $k = 5$ to $8$, and then $k = 9$ to $16$. This pattern means that $k = 4$, $8$, and $16$ are natural choices for the number of coarse-grained states, since any further increase (say from $k = 8$ to $9$) will yield a smaller improvement in explained variance. Similarly, the second reason for choosing $k=8$ comes from studying the average distance between states within a cluster, which is known as the dispersion [Fig. \ref{num_states}(c)]. Intuitively, a coarse-grained description with low dispersion provides a good fit of the observed data. Similar to the explained variance, we find that the decrease in dispersion from $k-1$ to $k$ clusters is nearly constant for $k = 3$ to $4$, then $k = 5$ to $8$, and then $k = 9$ to $16$, once again suggesting that $k = 4$, $8$, and $16$ are natural choices for the number of clusters. Together, these results demonstrate that the coarse-grained description with $k=8$ states provides a good fit to the neural time-series data while still allowing for an accurate estimate of the entropy production in each task.

\section{Flux networks: Visualizing fluxes between coarse-grained states}
\label{flux_nets}

\begin{figure*}
\centering
\includegraphics[width = .95\textwidth]{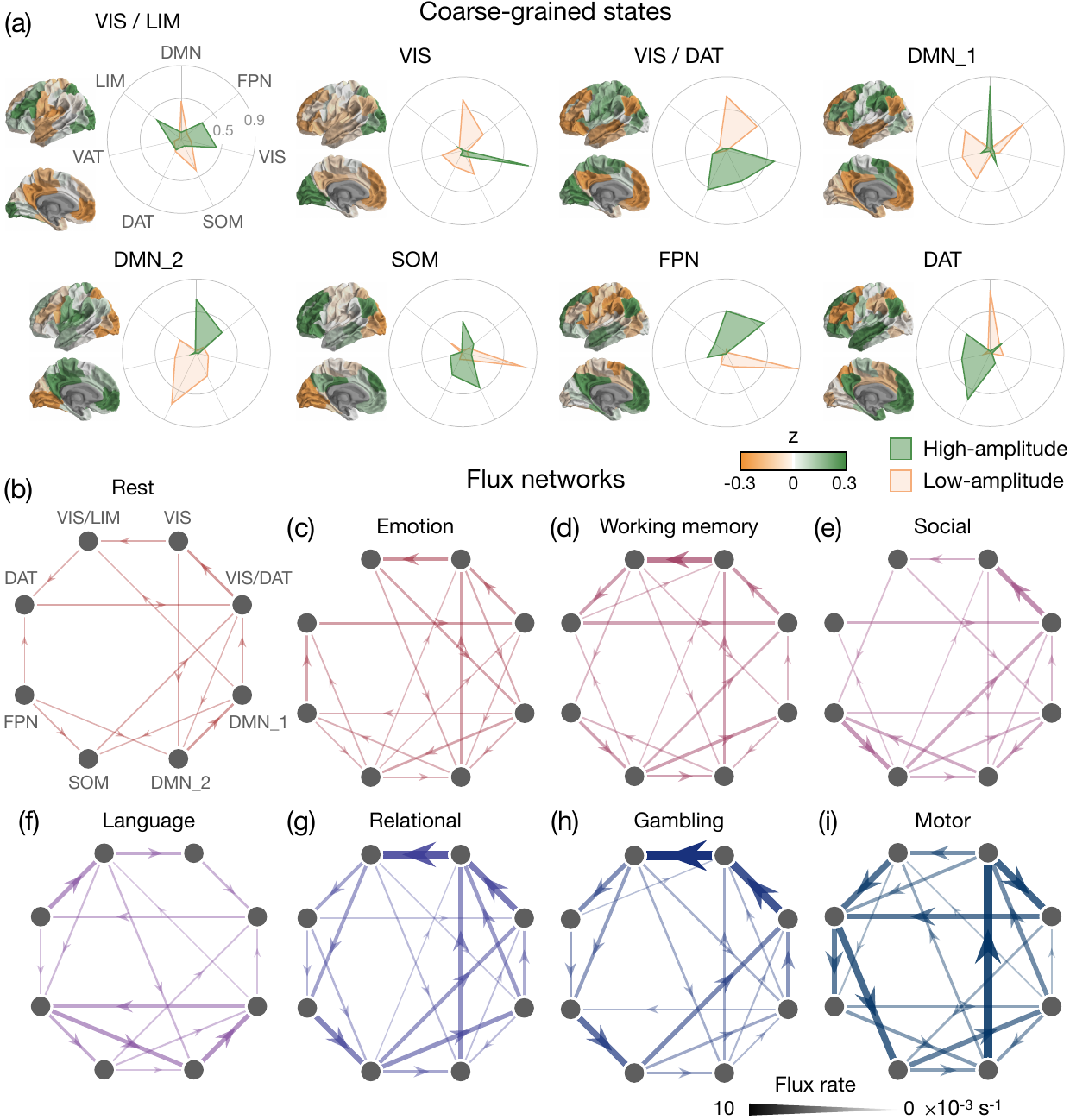} \\
\caption{\label{fluxNets} Flux networks reveal violations of detailed balance unique to each cognitive task. (a) Coarse-grained neural states calculated using hierarchical clustering ($k=8$), with surface plots indicating the z-scored activation of different brain regions. For each state, we calculate the cosine similarity between its high-amplitude (green) and low-amplitude (orange) components and seven pre-defined neural systems \cite{Thomas-01}: default mode (DMN),  frontoparietal (FPN), visual (VIS), somatomotor (SOM), dorsal attention (DAT), ventral attention (VAT), and limbic (LIM). We label each state according to its largest high-amplitude cosine similarities. (b-i) Flux networks illustrating the fluxes between the eight coarse-grained states at rest (b) and during seven cognitive tasks: emotional processing (c), working memory (d), social inference (e), language processing (f), relational matching (g), gambling (h), and motor execution (i). Edge weights indicate flux rates, and fluxes are only included if they are significant relative to the noise floor induced by finite data (one-sided \textit{t}-test, $p < 0.001$).}
\end{figure*}

In Fig. \ref{entProd}, we demonstrate that the brain has the capacity to operate at a wide range of distances from detailed balance. We did so by estimating the entropy production of neural dynamics during different cognitive tasks. In addition to investigating the entropy production, one can also examine the specific neural processes underlying the violations of detailed balance, which are encoded in the fluxes between neural states.

\begin{figure*}
\centering
\includegraphics[width = \textwidth]{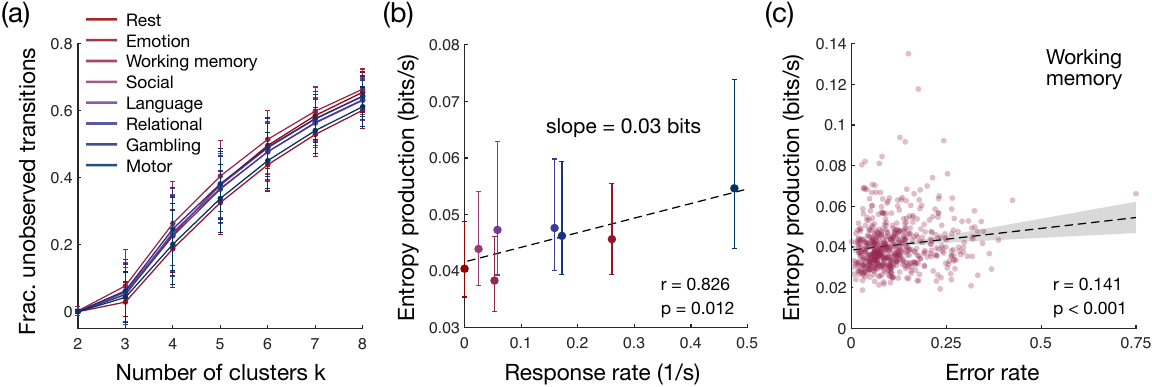} \\
\caption{\label{entProd_ind} Entropy production in individual subjects. (a) Fraction of the $k^2$ state transitions that remain unobserved after hierarchical clustering with $k$ clusters for each subject and each task. Data points and error bars represent two standard deviations over the 590 different subjects. (b) Entropy production of individual subjects increases with the rate of physical responses across the different task settings (Pearson correlation $r = 0.826$, $p = 0.012$). Data points and error bars represent medians and interquartile ranges over the 590 subjects, and the dashed line indicates linear best fit. (c) Entropy production increases with error rate in the working memory task (Pearson correlation $r = 0.141$, $p < 0.001$). We confirm that relationship also holds after removing outliers (Pearson correlation $r = 0.133$, $p = 0.002$). Dashed line and shaded region indicate linear best fit and 95\% confidence interval. To account for finite-data errors, all data points in all panels are averaged over 100 bootstrap samples for each subject and each task.}
\end{figure*}

We find that each of the $k=8$ coarse-grained states corresponds to high-amplitude activity in one or two cognitive systems \cite{Thomas-01} [Fig. \ref{fluxNets}(a)]. For each task, we can visualize the pattern of fluxes as a network, with nodes representing the coarse-grained states and directed edges reflecting net fluxes between states [Fig. \ref{fluxNets}(b-i)]. These networks illustrate, for example, that the fluxes almost vanish during rest [Fig. \ref{fluxNets}(b)], thereby indicating that the brain nearly obeys detailed balance. Interestingly, in the emotion, working memory, social, relational, and gambling tasks [Fig. \ref{fluxNets}(c-e,g,h)] -- all of which involve visual stimuli -- the strongest fluxes connect visual (VIS) states. By contrast, these fluxes are weak in the language task [Fig. \ref{fluxNets}(f)], which only involves auditory stimuli. Finally, in the motor task, wherein subjects are prompted to execute physical movements, the dorsal attention (DAT) state mediates fluxes between disparate parts of the network [Fig. \ref{fluxNets}(i)], perhaps reflecting the role of the DAT system in directing goal-oriented attention \cite{Fox-01, Vossel-01}. In this way, broken detailed balance in the brain is not driven by a single subsystem, but rather involves different combinations of subsystems depending on the specific task being performed. Examining the structural properties and cognitive neuroscientific interpretations of these flux networks is an important direction for future studies.

\section{Entropy production in individual humans}
\label{individual}

Throughout the main text, we combine the neural dynamics of all subjects in order to increase the statistical power of our analyses. However, it is also interesting to investigate violations of detailed balance in individual humans. The primary difficulty in doing so lies in estimating the transition probabilities $P_{ij}$ that are required to compute the entropy production [Eq. (\ref{S})]. As discussed in Appendix \ref{choosing_k}, concatenating the neural time-series across subjects allows us to estimate the transition probabilities using $k=8$ coarse-grained states (Fig. \ref{num_states}). By contrast, when analyzing individual subjects, the largest number of states for which we observe every transition at least once in each task is $k=3$ [Fig. \ref{entProd_ind}(a)].

Performing hierarchical clustering with $k=3$ clusters, we estimate the entropy production for each of the 590 subjects during each task (and rest). As in the main text (Fig. \ref{entProd}), we then investigate the dependence of entropy production on physical and cognitive exertion. Across all tasks, we find that the entropy production of neural dynamics increases significantly with the rate of motor responses [Fig. \ref{entProd_ind}(b)]. This result confirms that the population-level relationship between broken detailed balance and physical effort [Fig. \ref{entProd}(b)] extends to the scale of individual humans.

To examine the relationship between entropy production and cognitive demand, we once again focus on the working memory task. At the population level, we found that the high cognitive load condition induces a two-fold increase in entropy production over the low load condition [Fig. \ref{entProd}(c)]. However, performing the same analysis on individual subjects is infeasible, since it requires estimating the entropy production for each subject on only a fraction of the working memory data. Instead, as a proxy for cognitive effort, we can examine the rate at which subjects make errors. Indeed, across subjects, we find that entropy production increases significantly with error rate [Fig. \ref{entProd_ind}(c)], confirming that the association between broken detailed balance and cognitive effort persists at the individual level. Together, these results indicate that, even for individual humans, violations of detailed balance grow with physical exertion and cognitive demand.

\begin{figure*}
\centering
\includegraphics[width = .65\textwidth]{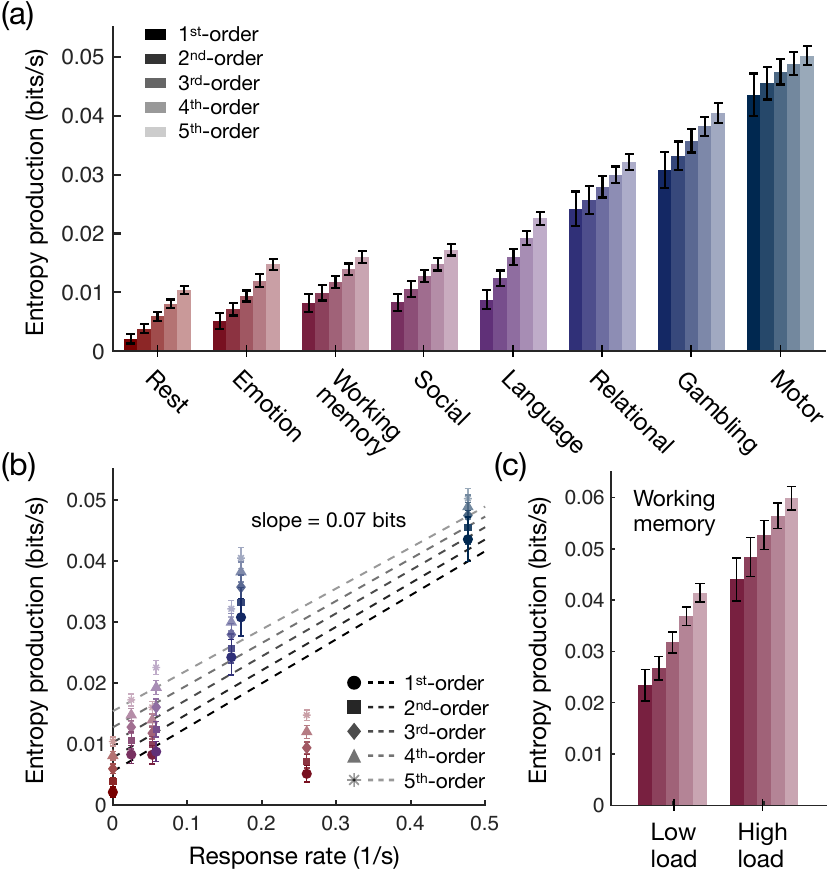} \\
\caption{\label{Markov} Higher-order approximations of entropy production in the brain. (a) Entropy productions of orders $\ell = 1,2,3,4,5$ computed at rest and during seven cognitive tasks. All estimates are based on the same coarse-grained states, computed using hierarchical clustering with $k=8$ clusters. (b) Entropy production estimates as a function of response rate for the tasks listed in panel (a). Across all orders $\ell=1,2,3,4,5$, each response induces an identical $0.07$ bits of produced entropy (within errors, $p < 0.05$). (c) Entropy production estimates for low cognitive load and high cognitive load conditions in the working memory task, where low and high loads represent 0-back and 2-back conditions, respectively. In all panels, data points of increasing brightness indicate entropy production estimates of increasing order, and error bars reflect-two standard-deviation confidence intervals that arise due to finite data (see Appendix \ref{methods}).}
\end{figure*}

\section{Testing the Markov assumption}
\label{assumption}

Thus far, we have employed a definition of entropy production [Eq. (\ref{S})] that relies on the assumption that the time-series is Markovian; that is, that the state $x_t$ of the system at time $t$ depends only on the previous state $x_{t-1}$ at time $t-1$. For real time-series data, however, the dynamics may not be Markovian, and Eq. (\ref{S}) is not exact. In general, the entropy production (per trial) is given by \cite{Roldan-01, Roldan-02}
\begin{equation}
\label{S2}
\dot{S} = \lim_{t \rightarrow \infty} \frac{1}{t} \sum_{i_1, \hdots, i_{t+1}} P_{i_1, \hdots, i_{t+1}} \log \frac{P_{i_1, \hdots, i_{t+1}}}{P_{i_{t+1}, \hdots, i_1}},
\end{equation}
where $P_{i_1, \hdots, i_{t + 1}}$ is the probability of observing the sequence of states $i_1,\hdots,i_{t+1}$. If the dynamics are Markovian of order $\ell$, then Eq. \ref{S2} is equivalent to
\begin{equation}
\label{S3}
\dot{S} = \frac{1}{\ell}\sum_{i_1, \hdots, i_{\ell+1}} P_{i_1, \hdots, i_{\ell+1}} \log \frac{P_{i_1, \hdots, i_{\ell+1}}}{P_{i_{\ell+1}, \hdots, i_1}}.
\end{equation}
For example, if $\ell = 1$ we recover the Markov approximation in Eq. (\ref{S}). In general, computing the $\ell^{\text{th}}$-order entropy production for a system with $k$ states requires estimating the probabilities of all $k^{\ell+1}$ sequences of length $\ell+1$. Thus, the number of independent statistics that need to be estimated grows exponentially with the order $\ell$, often making it infeasible to estimate the entropy production beyond order $\ell = 1$.

Despite the abovementioned limitations, here we estimate the entropy production of neural dynamics up to order $\ell = 5$. In doing so, we demonstrate that our main results (Fig. \ref{entProd}) do not depend qualitatively on the Markov approximation in Eq. (\ref{S}). Just as we did under the Markov approximation (Fig. \ref{entProd}), we cluster the neural data using $k=8$ coarse-grained states. Given that we are now required to estimate $k^{\ell+1}$ different probabilities (a number that grows up to $2.6\times 10^5$ for $\ell = 5$) rather than just $k^2 = 64$ probabilities, there are inevitably entries in the sum in Eq. (\ref{S3}) that are infinite (i.e., those corresponding to reverse-time sequences $i_{\ell+1},\hdots, i_1$ that are not observed in the time-series). As is common \cite{Roldan-01, Roldan-02}, we set these terms to zero.

We find that all of the higher-order approximations studied ($\ell = 2,3,4,5$) yield exactly the same hierarchy of entropy productions across task settings [Fig. \ref{Markov}(a)] as the first-order approximation [Fig. \ref{entProd}(a)]. In particular, across all orders $\ell$, the neural dynamics produce less entropy during rest than during each of the cognitive tasks [Fig. \ref{Markov}(a)]. Moreover, the higher-order entropy productions remain significantly correlated with the frequency of physical responses in different tasks [Fig. \ref{Markov}(b)]. In fact, for all orders $\ell$ examined, each response induces an identical $0.07$ bits of produced entropy [within errors; Fig. \ref{Markov}(b)]. Finally, in the working memory task, the higher-order entropy productions remain larger in the high cognitive load condition than in the low-load condition [Fig. \ref{Markov}(c)]. Specifically, the neural dynamics produce an additional $0.02$ bits per second of entropy in the high-load condition compared to the low-load condition, a difference that is identical (within errors) across all of the Markov orders studied. Together, these results demonstrate that the central conclusions of the main text generalize to higher-order Markov approximations.

\section{Varying the number of coarse-grained states}
\label{vary_k}

\begin{figure}[t!]
\centering
\includegraphics[width = .49\textwidth]{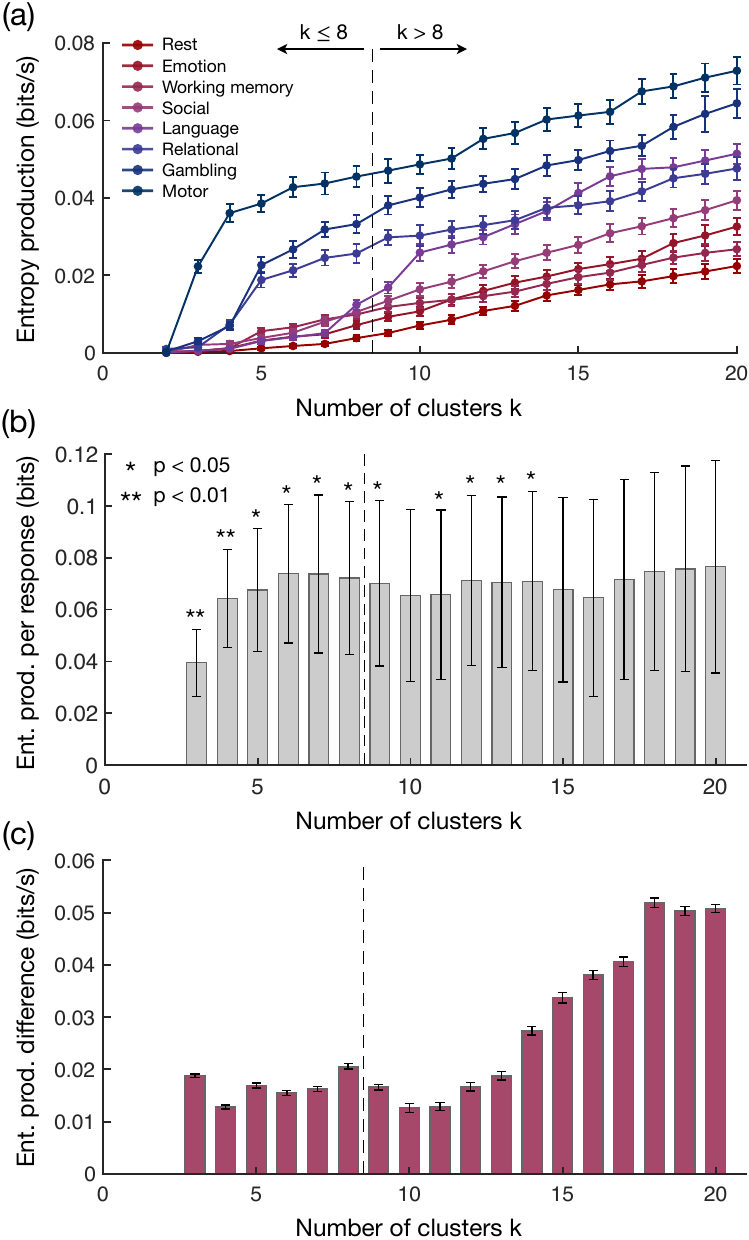} \\
\caption{\label{robust_k} Entropy production in the brain at different levels of coarse-graining. (a) Entropy production at rest and during seven cognitive tasks as a function of the number of clusters $k$ used in the hierarchical clustering. Error bars reflect two-standard-deviation confidence intervals that arise due to finite data (see Appendix \ref{methods}). (b) Slope of the linear relationship between entropy production and physical response rate across tasks for different numbers of clusters $k$. Error bars represent one-standard-deviation confidence intervals of the slope and asterisks indicate the significance of the Pearson correlation between entropy production and response rate. (c) Difference between the entropy production during high-load and low-load conditions of the working memory task as a function of the number of clusters $k$. Error bars represent two-standard-deviation confidence intervals that arise due to finite data (see Appendix \ref{methods}), and the entropy production difference is significant across all values of $k$ (one-sided \textit{t}-test, $p < 0.001$).}
\end{figure}

In Appendix \ref{choosing_k}, we presented methods for choosing the number of coarse-grained states $k$, concluding that $k=8$ is an appropriate choice for our neural data. However, it is important to check that the entropy production results from Fig. \ref{entProd} do not vary significantly with our choice of $k$. In Fig. \ref{robust_k}(a), we plot the estimated entropy production for each task setting (including rest) as a function of the number of coarse-grained states $k$. We find that the tasks maintain approximately the same ordering across all choices of $k$ considered, with the brain producing the least entropy during rest, the most entropy during the motor task, and the second most entropy during the gambling task. Furthermore, we find that the correlation between entropy production and physical response rate [Fig. \ref{entProd}(b)] remains significant for all $k \le 8$ [that is, for all choices of $k$ for which we observe all transitions at least once in each task; Fig. \ref{num_states}(a)] as well as $k = 9$, $11$, $12$, $13$, and $14$ [Fig. \ref{robust_k}(b)]. We remark that we do not study the case $k=2$ because the entropy production is zero by definition for steady-state systems with two states [Fig. \ref{robust_k}(a)]. Finally, we confirm that the entropy production is significantly larger during high-cognitive-load conditions than low-cognitive-load conditions in the working memory task [Fig. \ref{entProd}(c)] for all choices of $k$ considered [Fig. \ref{robust_k}(c)]. Together, these results demonstrate that the relationships between entropy production and physical and cognitive effort are robust to reasonable variation in the number of coarse-grained states $k$.

\section{Robustness to head motion and signal variance}
\label{robust}

\begin{figure}
\centering
\includegraphics[width = .4\textwidth]{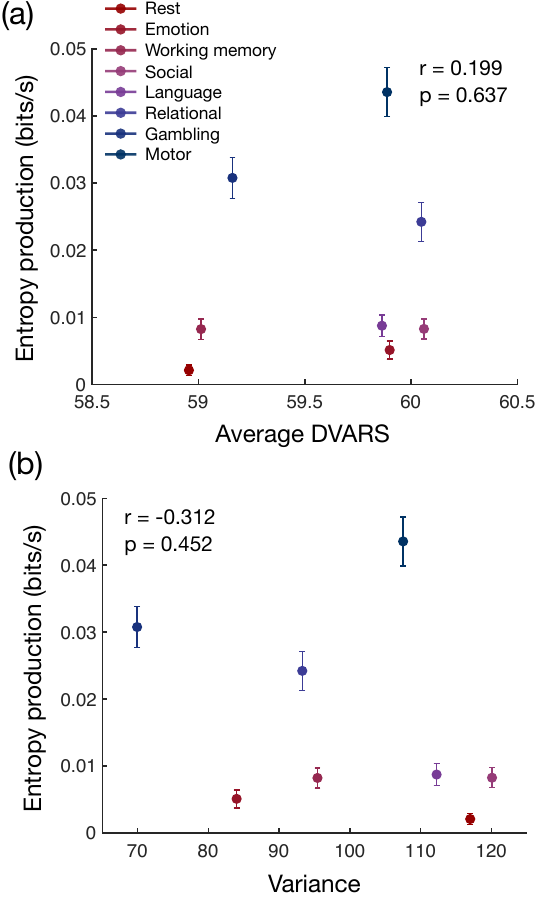} \\
\caption{\label{DVARS} Entropy production in the brain cannot be explained by head movement nor signal variance. Entropy production versus the average DVARS (a) and the variance of the neural time-series (b) at rest and during seven cognitive tasks. Across both panels, entropy productions are estimated using hierarchical clustering with $k=8$ clusters and are divided by the fMRI repetition time $\Delta t = 0.72$ s to compute entropy production rates. Error bars reflect two-standard-deviation confidence intervals that arise due to finite data (see Appendix \ref{methods}).}
\end{figure}

Here, we show that the effects of physical and cognitive effort on entropy production cannot be explained by head movement within the scanner (a common confound in fMRI studies \cite{Friston-05}) nor variance in the neural time-series. To quantify head movement, for each time point in every time-series, we compute the spatial standard deviation of the difference between the current image and the previous image. This quantity, known as DVARS, is a common measure of head movement in fMRI data \cite{Power-01}. Importantly, we find that entropy production is not significantly correlated with the average DVARS within each task [Fig. \ref{DVARS}(a)], thereby demonstrating that the relationship between entropy production and physical response rate is not simply due to the confound of subject head movement within the scanner. Additionally, we find that entropy production is not significantly correlated with the variance of the neural data within each task [Fig. \ref{DVARS}(b)]. This final result establishes that our entropy production estimates are not simply driven by variations in the amount of noise in the neural data across different tasks.

\section{Data processing}
\label{processing}

The resting, emotional processing, working memory, social inference, language processing, relational matching, gambling, and motor execution fMRI scans are from the S1200 Human Connectome Project release \cite{Barch-01, Van-01}. Brains were normalized to fslr32k via the MSM-AII registration with 100 regions \cite{Schaefer-01}. CompCor, with five principal components from the ventricles and white matter masks, was used to regress out nuisance signals from the time series. Additionally, the 12 detrended motion estimates provided by the Human Connectome Project were regressed out from the regional time series. The mean global signal was removed and then time series were band-pass filtered from 0.009 to 0.08 Hz. Then, frames with greater than 0.2 mm frame-wise displacement or a derivative root mean square (DVARS) above 75 were removed as outliers. We filtered out sessions composed of greater than 50 percent outlier frames, and we only analyzed data from subjects that had all scans remaining after filtering, leaving 590 individuals. The processing pipeline used here has previously been suggested to be ideal for removing false relations between neural dynamics and behavior \cite{Siegel-01}. Finally, for each subject and each scan, we only analyze the first 176 time points, corresponding to the length of the shortest task (emotional processing); this truncation controls for the possibility of data size affecting comparisons across tasks.

\newpage
\clearpage

\bibliography{DetailedBalanceBib}

\end{document}